\documentclass[preprint,showpacs,preprintnumbers,amsmath,amssymb]{revtex4}

\usepackage{color}

\usepackage{graphicx}
\usepackage{dcolumn}
\usepackage{bm}

\definecolor{magenta}{rgb}{1.0,0.0,1.0}

\newcommand{\blue}{\textcolor{blue}}
\newcommand{\red}{\textcolor{red}}
\newcommand{\green}{\textcolor{green}}
\newcommand{\magenta}{\textcolor{magenta}}

\newcommand{\be}{\begin{equation}}
\newcommand{\ee}{\end{equation}}

\newcommand{\bea}{\begin{eqnarray}}
\newcommand{\eea}{\end{eqnarray}}

\begin{document}

\preprint{CERN-PH-TH/2004-084}  

\preprint{IFT-UAM-CSIC-04-19}

\vspace{3cm}

\title{Regular compactifications and Higgs model vortices}

\author{B.~de Carlos}
\email{B.de-Carlos@sussex.ac.uk}
\affiliation{Department of Physics and Astronomy, University of Sussex, \\ Falmer, Brighton BN1 
9QJ, UK }
 \altaffiliation[Also at ]{Department of Physics CERN, Theory Division, 1211 Geneva 23, Switzerland}
 \author{J.M.~Moreno}%
 \email{jesus.moreno@uam.es}
\affiliation{Instituto de F\'{\i}sica Te\'orica \\ C-XVI, UAM \\ 28049 Madrid, Spain}%

\date{\today}

\vspace{3cm}

\begin{abstract}
We present full numerical solutions to the system of a global string embedded in a six-dimensional
space time. The solutions are regular everywhere and do confine gravity in our four-dimensional
world. They depend on the value of the (negative) cosmological constant in the bulk and on the parameters of
the Higgs potential, and we perform a systematic study to determine their allowed values.
We also comment on the relation of our results with previous studies on
the same subject and on their phenomenological viability. 
\end{abstract}

\pacs{04.50.+h,  11.10.Kk }
\maketitle

\section{\label{sec:level1}Introduction}

Field theory constructions in which our four-dimensional world consists of a topological defect 
embedded in a higher dimensional space-time are more and more frequent these days, although
they were originally proposed twenty years ago \cite{Akama:jy,Rubakov:1983bz,Visser:qm}.
Such defect could either be a domain wall (if the total number of dimensions is five), a string (six),
a monopole (seven) or an instanton (eight). 

The gravitational field of these topological defects becomes relevant in this context. In $D=4$ 
that of the  domain wall has been thoroughly studied \cite{Vilenkin:hy,Ipser:1983db,Cvetic:1996vr,
Gibbons:in,Cvetic:1993xe,Bonjour:1999kz}. Whereas it seems to be  non-static from the perspective of an observer 
on the wall, it is actually the domain wall which is non-static, moving in a static
Minkowski space-time. The gravitational field of the global monopole is as well static and well-defined
\cite{Barriola:hx}, and it is only the global string which happens to give rise to a static and singular
metric outside the core of the defect \cite{Cohen:1988sg}. It was later realised that this singularity
could be cured by adding time dependence to the metric \cite{Gregory:1996dd}.

Coming back to higher dimensions, it is then mandatory to explore the gravitational field generated 
by these defects if one is willing to build up realistic models in which our four dimensional world
is to become one of them. Following the work of Randall and Sundrum \cite{Randall:1999ee},
the domain wall in $D=5$ has been extensively studied, and we have nothing new to add to it.
Also, the generalization of the global monopole to higher dimensions, which results in a $D=7$
space-time, has been studied \cite{Olasagasti:2000gx}, resulting in, again, a well-defined static metric.
The issue then remained of whether the gravitational field generated by the global string would still result 
singular in its $D=6$ format.

This issue was first investigated by Cohen and Kaplan \cite{Cohen:1999ia}, who concluded that, 
similarly to the 
four-dimensional case, the metric around the global string is still singular.
Then Gregory \cite{Gregory:1999gv} argued that, again analogously to the $D=4$ case, adding time
dependence to the metric should cure that problem and that an equivalent procedure would be
to add to the static metric a negative cosmological constant in the bulk. Higher dimensional
extensions of these claims were presented in \cite{Gregory:2002tp}. In these last two articles
analytic arguments were given to support them, although  a full numerical solution was not
presented.

This is precisely the main aim of the present letter. We want to present numerical evidence of the 
existence of solutions that confine gravity around a global string in which core our four-dimensional
world exists. The geometry of the $D=6$ space-time is that of a cigar, and it behaves asymptotically 
as $AdS_5 \times S_1$. These solutions are numerically hard to obtain and rely on a very precise tuning
of the value of the (negative) cosmological constant in the bulk.

In the next section we describe the model we are going to work with, namely the system of a scalar
field together with gravity in $D=6$, and we write the set of equations to which their dynamics
reduce. In section~3 we show our solutions, explaining the technical involvement of finding them,
and also  their physical interpretation. We compare with related models already published in the literature and, 
in section~4, we conclude.

\section{The model}

A global string is a topologically non trivial solution of the following Lagrangian
\be
{\cal L} = \frac{1}{2}(\nabla_{A} \, \Phi )^\dagger \nabla^A \Phi 
- V(\Phi) 
\label{Lag_global}
\ee
where the potential has to exhibit a global U(1) symmetry. In particular, we shall study the 
so-called Mexican-hat potential, i.e.
\be
V(\Phi) = \lambda(\Phi^\dagger \Phi-v^2)^2 \;\;,
\label{higgs_pot}
\ee
Our goal is to find numerical solutions for the equations that govern the system formed by the
field $\phi$ when considered in the context of the six-dimensional geometry defined by 
\be
ds^2 = M^2(r) \eta_{\mu \nu} x^\mu x^\nu - dr^2 - L^2(r) d \theta^2 \ ,
\label{6metric}
\ee
where $(r,\theta)$ are the coordinates of the transverse space and $M(r)$, $L(r)$ are the warp factors.
 In particular we are looking for solutions that would result in gravity trapping in four space-time
 dimensions.

Our starting point will be the action for the $D=6$ system
\be 
S = -\int dx^6 \sqrt{-g} \left( \frac{R}{2 \kappa^2} + \Lambda - {\cal L} \right) \;\;, 
\label{action}
\ee
with $g$ the determinant of the metric, eq.~(\ref{6metric}), $\kappa^2 = 1/M_6^4$, and $M_6$ is the $D=6$ Planck mass.
$\Lambda$ is the bulk cosmological constant and ${\cal L}$ is given by eq.~(\ref{Lag_global}).

The Einstein equations for this system are given by
\be
R_{ab} - \frac{1}{2} g_{ab} R = \frac{1}{M_6^4} (\Lambda g_{ab} + T_{ab}) \;\;,
\label{Einstein}
\ee
where $a=1,\ldots,6$ and the equation of motion for the field $\Phi$ is given by
\be
g^{ab} \nabla_a \nabla_b \Phi + \frac{\partial V}{\partial \Phi^{\dagger}} = 0 \;\;.
\label{eom}
\ee
In order to simplify the presentation of results we shall  parametrize the scalar field as
\be 
\Phi = v F(r) {\rm e}^{iq \theta},
\ee
and we define $m(r) \equiv M'(r)/M(r)=d {\rm ln} M(r)/dr$. Our task is, therefore, to solve a set of second 
order differential equations for the variables $f(r)$, $L(r)$ and $m(r)$ for certain values of
the parameters $\Lambda$, $v$ and $q$. The actual equations, already assuming $M_6=1$, are given by
\bea
\frac{L''}{L} + 3 m' + 6 m^ 2 + 3 \frac{L'}{L}m & = & - \Lambda  -
\left( \frac{v^2F'^2}{2} + \frac{q^ 2 v^2F^2 }{2 L^ 2} + \lambda v^4(F^2-1)^2 \right)  \;\;,  \nonumber \\
4 m' + 10m^ 2 &=& - \Lambda - \left( \frac{v^2F'^2}{2} - \frac{q^ 2 v^2 F^2 }{2 L^ 2} + 
\lambda v^4 (F^2-1)^2 \right) \;\;,  \label{Einstein+eom} \\
2 \frac{L'}{L}m + 3 m^2  &= & - \Lambda - \frac{1}{2}
\left( -v^2 F'^2 + \frac{q^ 2 v^2 F^2 }{2 L^ 2} + \lambda v^4 (F^2-1)^2 \right) \;\;, \nonumber  \\
F'' + \left( 4 m +  \frac{L'}{L} \right) F' &=& 4 \lambda F v^2 (F^2-1) + \frac{q^2 F }{L^2}  \nonumber \;\;.
\eea
Therefore we have to solve three differential equations (the third one of the previous system is
a constraint) with a mixture of boundary conditions defined at the origin and at infinity. This is a
typical boundary value problem.

To be more precise, let us elaborate on the choice of boundary conditions: at the origin ($r \rightarrow 0$), we 
demand a regular geometry at the core of the string and the absence of deficit angle in
our solution. This translates into
\be
m(0) = 0,  \; \;   L(0) = 0 \: \:,  L'(0) = 1 \;\;.
\ee
Moreover, a local analysis by power series shows that, near the origin,
\bea
F(r)  & = & f_1 r  \;\;, \nonumber \\
L(r)  & =   & r + l_3 r^3 \;\;, \\
m(r) & = & m_1 r \;\;. \nonumber
\label{origen}
\eea
Substituting these ansatze in the previous equations (\ref{Einstein+eom}) we get
\bea
m_1 & = & -\frac{1}{4}(\Lambda + V(0)) \;\;, \nonumber \\
l_3 & = & \frac{1}{12}(\Lambda + V(0)-2 f_1^ 2) \;\;.
\eea
$f_1$ remains a free (shooting) parameter.

Far away from the core, at $r \rightarrow \infty$, we demand that all three functions 
in eqs.~(\ref{origen}) go to constants. The metric in this region is then
assumed to be cigar-like

\be
ds^2_\infty = e^{2 m_\infty r} \eta_{\mu \nu} x^\mu x^\nu - dr^2 - 
L^2_\infty d \theta^2 \ .
\label{infinite-metric}
\ee
Substituting again in eqs.~(\ref{Einstein+eom}), and reinstating factors of $M_6$ where necessary, we get
\be
m_\infty = \pm \frac{\sqrt{-( \Lambda + V(f_\infty))}}{ 2 \sqrt{2} M_6^2} \;\;.
\label{minf}
\ee
For phenomenological reasons, i.e. in order to have gravity trapping in $D=4$,  
we are looking for solutions that
correspond to $m_{\infty} < 0$. The second warp factor is given by
\be
L_\infty =\frac{ f_\infty }{ 2  M_6^2 |m_\infty|} \;\;,
\label{Linf}
\ee
whereas $f_{\infty}$ is obtained by solving the equation
\be
 \Lambda + V(f_\infty) - 2 f_\infty V'(f_\infty) = 0 \;\;.
\label{finf}
\ee
The first interesting conclusion to be drawn is that the solution of the previous equation will
{\em not} correspond to the field settling at the minimum of its potential, i.e. $V'(f_\infty)=0$. The presence 
of the negative cosmological constant, in other words, the interplay of the scalar field with gravity 
induces the field to settle just before reaching its minimum, as we will see next.

\section{Results}

Once we have defined the system we want to work with, we can look for solutions that satisfy
our requirements. In order to perform a numerical analysis we have used a relaxation method
which would look for a solution once values for $\Lambda$, $q$ and $v$ were specified. This means that we replace 
the system given by eq.~(\ref{Einstein+eom}) by a set of finite-difference equations (FDEs) on a mesh of points
that spans from zero to a sufficiently large value for $r$. Given that we are starting off with a system of 5
coupled first-order equations (remember that, in eq.~(\ref{Einstein+eom}), one equation is a constraint and another 
is already first order) represented by FDEs on a mesh of $M$ points, the solution consists of values for $5\times M$
variables. The relaxation method determines the solution by starting with a guess and improving it iteratively. 
As the iterations improve the solution, the result is said to relax to the true one.

From now on we will focus on the case $q=1$. Then, for every value of $v$ there is a unique value of
$\Lambda$ (which we shall call $\Lambda_{\rm c}$) which gives us a regular 
(with no deficit of solid angle) solution everywhere.
This was already pointed out by Ruth Gregory \cite{Gregory:1999gv} and here we present the first numerical 
evidence of this statement.

The way in which we determine $\Lambda_c$ is
as follows: in our relaxation code we give as boundary conditions the values 
of $F(r)$,
$L(r)$ and $m(r)$ at the origin, and those of $F'(r)$ and $L'(r)$ at infinity, 
which corresponds to a typical boundary value problem. This means that we do 
not know a priori the value of $L'(0)$, which determines whether or not there 
is a deficit of solid angle. Therefore, for
every value of $v$ we try different values of $\Lambda$ until we find a 
solution with
$L'(0)=1$.

In order to simplify the description of the results, we shall use adimensional variables to draw the plots, which 
requires the following reparametrization.
\bea
x & = & \sqrt{\lambda} v r \;\;, \nonumber  \\
f(x) & = & F(r)  \;\;, \label{reparam} \\
L(x) & = & \sqrt{\lambda} v L(r) \;\;, \nonumber \\
m(x) & = & m(r)/(\sqrt{\lambda}v) \;\;. \nonumber 
\eea
This already shows that the results are independent of the value of $\lambda$.

\begin{figure}[htbp]
\begin{center}
\includegraphics*[width=10cm, draft=false]{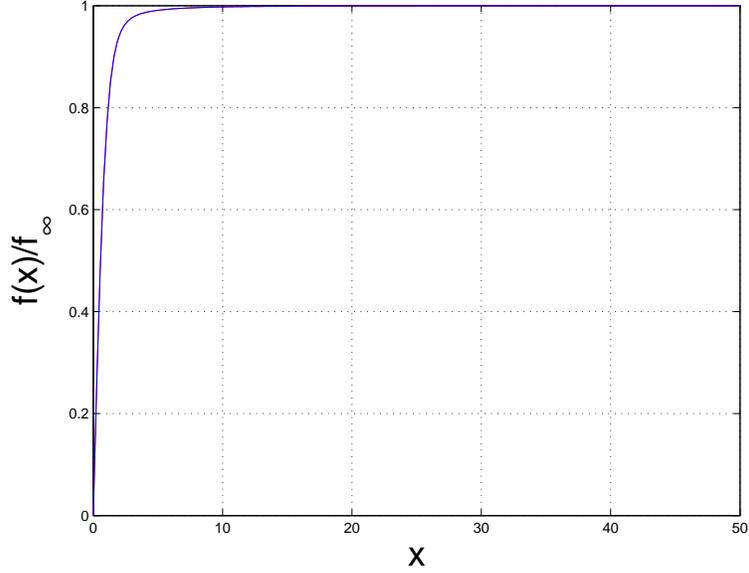}
\caption{Plot of $f(x)/f_{\infty}$ as a function of $x$ for $v=1$, $\Lambda_{c}=-0.0671$.}
\label{fig1}
\end{center}
\end{figure}
\begin{figure}[htbp]
\begin{center}
\includegraphics*[width=10cm, draft=false]{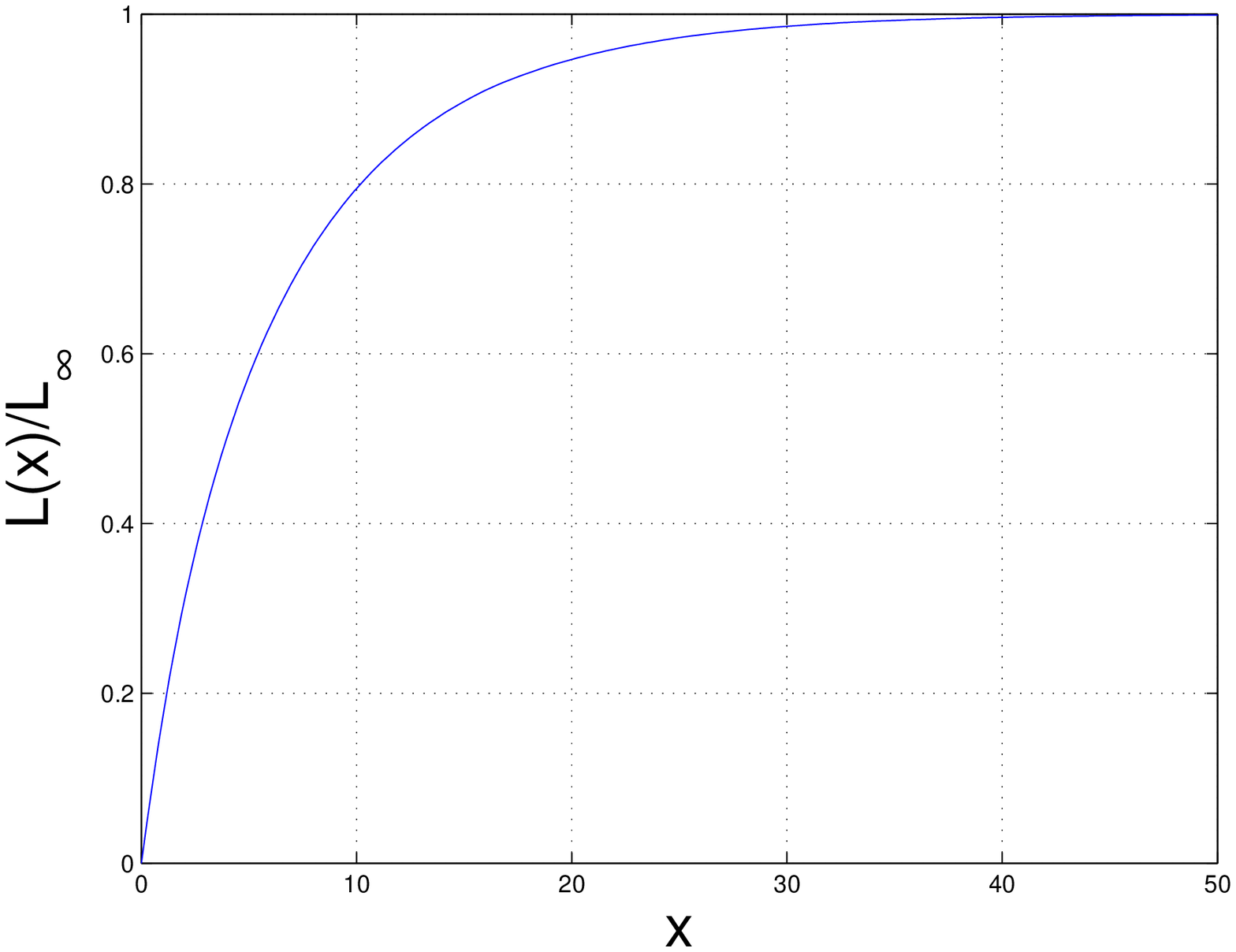}
\caption{Plot of $L(x)/L_{\infty}$ as a function of $x$ for $v=1$, $\Lambda_{c}=-0.0671$.}
\label{fig2}
\end{center}
\end{figure}
\begin{figure}[htbp]
\begin{center}
\includegraphics*[width=10cm, draft=false]{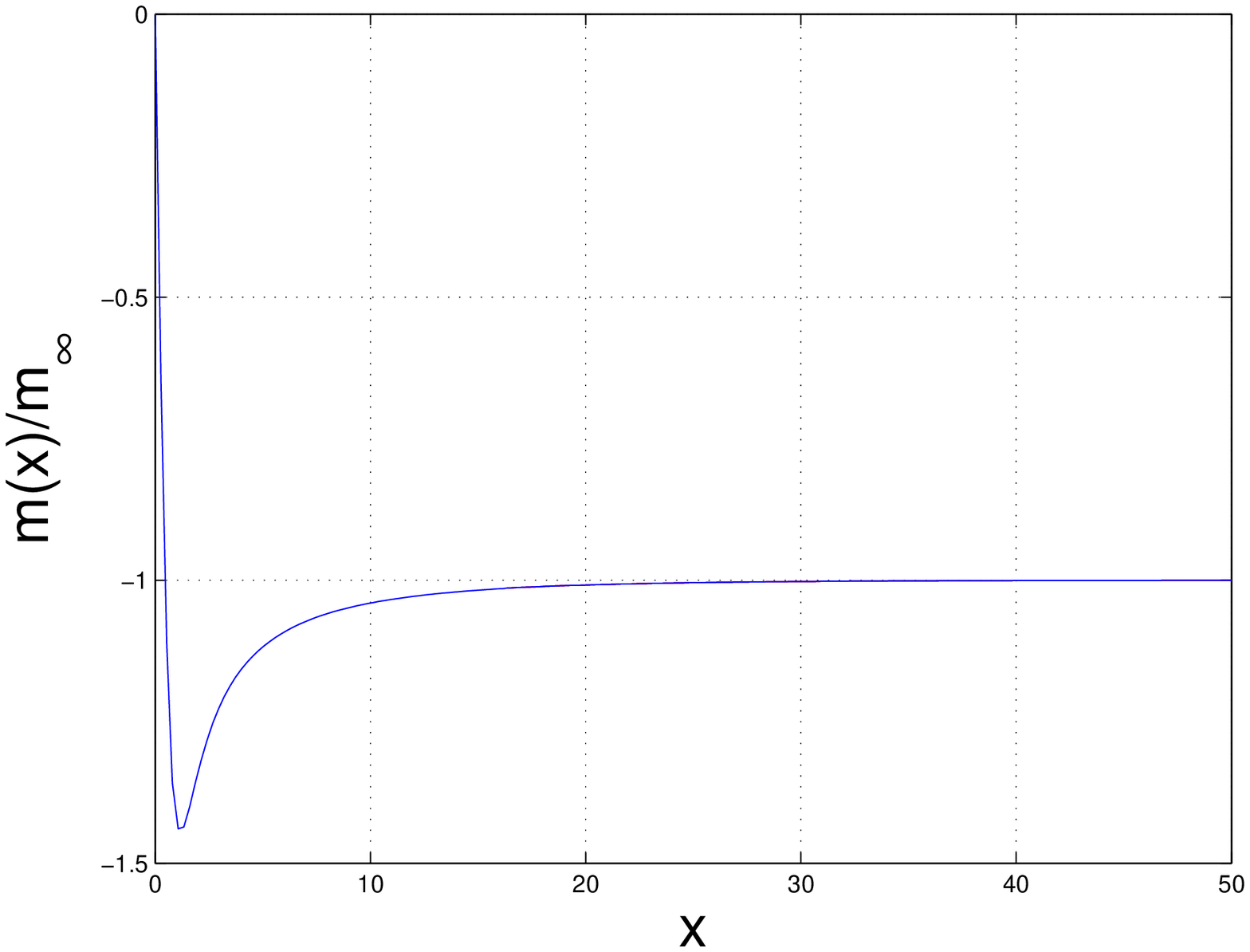}
\caption{Plot of $m(x)/m_{\infty}$ as a function of $x$ for $v=1$, $\Lambda_{c}=-0.0671$.}
\label{fig3}
\end{center}
\end{figure}
In figure~1 we show the evolution of the scalar field, represented by $f(x)$ normalized 
to its value at infinity, which is given by $f_\infty=0.99577$, i.e. the field gets stabilized 
just before its minimum located at $v=1$. As for the two warp factors, $L(x)$ and $m(x)$,
shown in figures~2 and 3 respectively, we can see how both go to constants at infinity.
The associated geometry of the space-time is, therefore, that of a {\em cigar},
which goes to $AdS_5 \times S_1$ asymptotically.

We have performed a systematic analysis of the parameter space, by varying the value
of $v$ and adjusting the corresponding one of $\Lambda_{\rm c}$. The results are
shown in figure~4, where we plot the warp factor $m(x)$ as a function of $x$ for $v=1.4,1,0.7,0.6$.
As we can see, the smaller $v$ is the longer it takes the gravity fields to settle at their asymptotic values.
This is just reflecting the fact that, for smaller $v$, the scalar field tends to settle closer and closer to its
minimum, playing less of a role in the stabilization of the warp factors. In other words the
problem becomes a typical two-scale one, where the scalar field quickly converges whereas the
other two slowly flow to their asymptotic values. Therefore, the numerical involvement of the
problem increases as we decrease the value of $v$.
\begin{figure}[htbp]
\begin{center}
\includegraphics*[width=10cm, draft=false]{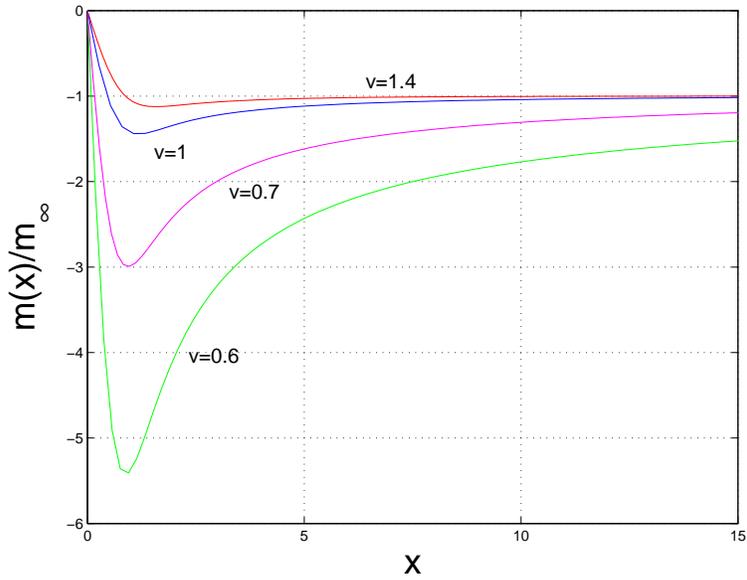}
\caption{Plot of $m(x)/m_{\infty}$ as a function of $x$ 
for \red{ $v=1.4$}, \blue{$v=1$}, 
\magenta{ $v=0.7$} and \green { $v=0.6$} }
\label{fig4}
\end{center}
\end{figure}

To further comment on numerical issues, let us describe how difficult it 
is to obtain these solutions.
As mentioned before, for a given model, there exists a regular solution just for a unique value of
$\Lambda$.  This was thoroughly explained in Ref.~\cite{Gregory:1999gv}, and here we shall merely repeat the 
main arguments and give a numerical proof. Essentially it is fair to study this system in
its asymptotic region by assuming the scalar field $f(r)$ to be at its minimum and working with
just the two warp factors $L(r)$ and $m(r)$. Then we can define the following variables
\bea
x & = & \frac{1}{-2 \Lambda}  \left( 3m+\frac{L'}{L} \right) \;\;, \nonumber \\
& & \\
y & = & \frac{1}{-2 \Lambda} \frac{L'}{L}  \;\;, \nonumber
\label{var_auto}
\eea
which, together with the independent variable $\rho=x \sqrt{-2 \Lambda}$, define the following
autonomous dynamical system,
\bea
x' & = & \frac{xy}{3} - \frac{4}{3} y^2 \;\;, \nonumber \\
& & \label{autu} \\
y' & = & \frac{4}{3} x(x-y) - y^2 - \frac{3}{4} \;\;. \nonumber
\eea
The primes here mean derivatives with respect to $\rho$.

In order to understand the structure of these numerical solutions, we must calculate the
critical points of this system. Those are given by
\bea
c_1^{\pm} & = & \left( \pm \frac{3}{4}, 0 \right) \hspace{5cm}  {\rm saddle} \nonumber \\
& & \label{fixed} \\
c_2^{\pm} & = & \pm \left( \sqrt{\frac{4}{5}}, \frac{1}{\sqrt{4\times5}} \right) \hspace{3cm}
{\rm attractor}/{\rm repellor}  \nonumber
\eea
The solutions we have found, which we have shown in previous plots, 
correspond to flowing towards $c_1^-$  
and it is now easy to understand why they are so hard to obtain: 
this is a saddle point, which is next to a 
repellor, given by $c_2^-$ (which, by the way, would 
be the critical point describing a geometry of the type 
$AdS_6$). 
Essentially only one trajectory, 
the one corresponding to $\Lambda_{\rm c}$,  ends up in $c_1^-$ which 
can be matched to a regular solution near the core of the string. 
Solutions ending up in $c_1^+$ would correspond to 
a four dimensional metric that would blow-up.

\begin{figure}[htbp]
\begin{center}
\includegraphics*[width=10cm, draft=false]{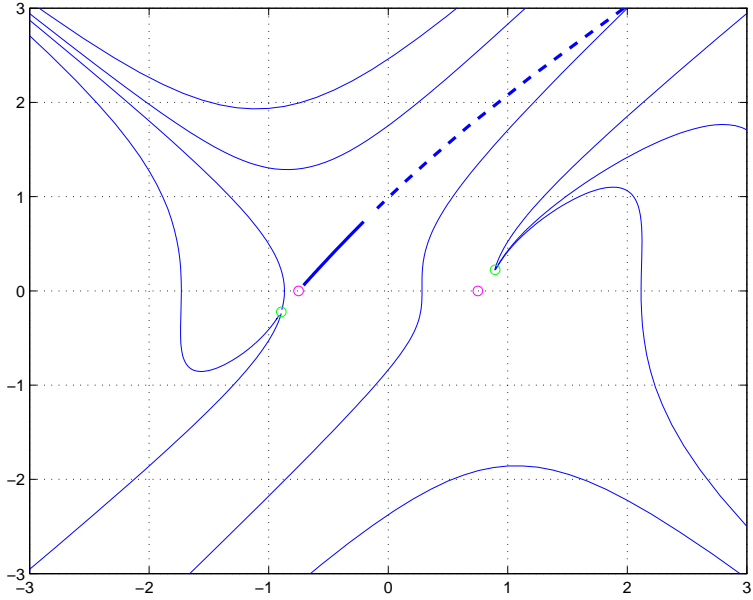}
\caption{Plot of the $(x,y)$ phase space, for $v=1$.}
\label{fig5}
\end{center}
\end{figure}
For completeness we give a plot of the phase space for this autonomous system in figure~5,
where we have also inserted the solution we have found (with a thick line). As we can see, far enough from 
the saddle point $c_1^-$, the scalar field $F(r)$ still plays a role in the evolution of $x$ and $y$
and, therefore, the inserted trajectory does not fit too well within those obtained from solving the system 
(\ref{autu}). We have highlighted this by using dashed lines in that part of our solution. As the field
$F(r)$ plays less of a role, the trajectory starts to fit 'naturally' into the phase space drawn, and
that is represented by the continuous thick line.

In fact, one can even go further and obtain an analytic expression for the two warp factors $L(r)$ and $m(r)$ 
as they approach their asymptotic limit. This is done by rewriting the autonomous system Eq.~(\ref{autu})
in the linear approximation around $c_1^-$, and solving for the two functions. The final  
result is that
\bea
{\rm ln} \left( \frac{L(r)}{L_\infty} \right) & = & - A {\rm e}^{2 (1-\sqrt{3}) m_{\infty} r} \;\;, \nonumber \\
& & \\
\frac{m(r)}{m_{\infty}} & = & -\left(1- (1-2/\sqrt{3}) A {\rm e}^{2(1-\sqrt{3}) m_{\infty} r} \right) \;\;. \nonumber
\eea
The $A$ factor, which determines the normalization of the curves, can then be extracted from
our numerical results. We have checked that the results we get for $A$ from the two curves, i.e. $L(r)$ 
and $m(r)$, are compatible with each other and that the fit to both functions in the asymptotic region 
is extremely accurate.

Next, we would like to discuss the role played by the cosmological constant in the bulk, 
$\Lambda$. As it was mentioned at the beginning of this section, we have numerically checked
that, for every value of $v$ there is a unique value of $\Lambda$, which we denote as $\Lambda_c$,
that gives us a regular solution everywhere. We have explored the parameter space defined by
$\Lambda_c/\lambda$ and $v$, and we have compared our results with 
existing ones in the literature.
\begin{figure}[htbp]
\begin{center}
\includegraphics*[width=10cm, draft=false]{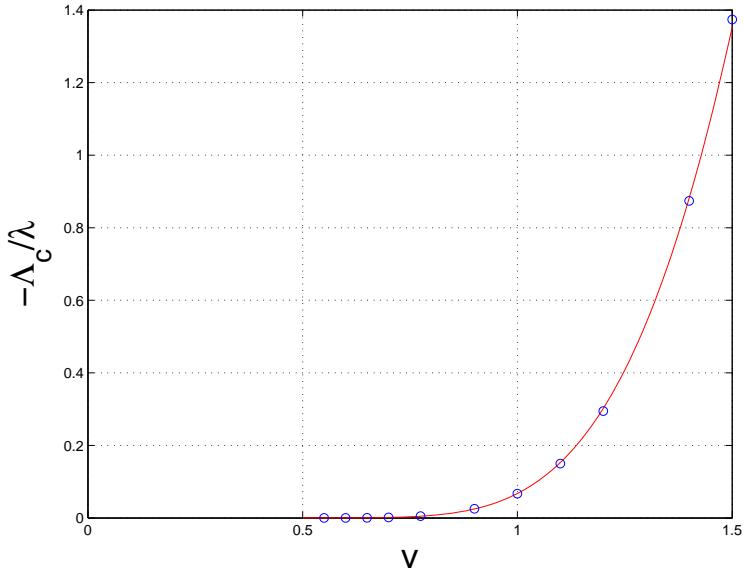}
\caption{Plot of $-\Lambda_c/\lambda$ as a function of  $v$, for small values of the latter. 
We have superimposed to the data the best fit we could find, given by eq.~(\ref{smallv}) with $a=6.3$, 
$b=9.0$.}
\label{fig6}
\end{center}
\end{figure}
For small values of $v$, see figure~6, we find a very good linear fit to the quantity 
$v {\rm log}(-\Lambda_c/\lambda)$, which means that, in that region
\be
\frac{\Lambda_c}{\lambda} = -{\rm e}^{a-b/v} \;\;.
\label{smallv}
\ee
We have compared these results to the estimate given in Ref.~\cite{Gregory:1999gv}, i.e.
$\Lambda \sim \epsilon {\rm e}^{-1/\epsilon}$, where $\epsilon=v^2$. This estimate was made for
very small values of $v$ and assuming the constraint $\lambda v^2=1$. Taking these facts into
account,  we find remarkable similarity between that result and the numbers we obtain as, in our language, 
that estimate reads $\Lambda_c/\lambda \sim {\rm e}^{-1/v^2}$, to be compared to eq.~(\ref{smallv}). However, 
as mentioned above, numerical difficulties prevent us from exploring the
region of very small $v$ and, therefore, we cannot check the validity of this equation in that region.

Having already mentioned the numerical difficulties 
encountered when exploring the small $v$ region, which corresponds to the weak gravity limit,
we turned to explore large values for $v$. As it happens, there is an upper bound on the value of
$\Lambda_c$ if we want to obtain regular solutions that {\em trap} gravity. This is given by the
quantity $m_1/v^2$ vanishing, or, in other words, by
\be
1+ \frac{\Lambda_c}{\lambda v^4} \rightarrow  0 \;\;.
\label{superheavy}
\ee
We show this graphically in figure~7, where we plot $-\Lambda_c/(\lambda v^4)$ as a 
function of $v$, indicating the points that were calculated numerically. We can clearly see that we
are approaching the super heavy limit, given by eq.~(\ref{superheavy}), and we can calculate how
the different relevant quantities in our problem approach it. For example, the asymptotic value of
the scalar field, $f_{\infty}$, using eq.~(\ref{finf}), will be given by
\be
f_{\infty}^{sh} = \sqrt{\frac{6}{7}}  v \;\;,
\ee
which means that, in the limit where gravity is strongest, the field will be maximally displaced from the 
minimum of its potential, at a value given by
\be 
V(f_{\infty}^{sh}) = \lambda \frac{v^4}{49} \;\;.
\ee
The other two asymptotic quantities will be given by
\bea
m_{\infty}^{sh} & = & \pm \sqrt{6 \lambda} \frac{v^2}{7 M_6^2} \nonumber \\
& & \\
L_{\infty}^{sh} & = & \frac{\sqrt{7}}{2} \frac{1}{\sqrt{\lambda}v} \;\;. \nonumber
\eea
Note that, in terms of adimensional quantities, 
$m_{\infty}^{sh} = \sqrt{6}v/(7 M_6^2)$, and $L_{\infty}^{sh} = \sqrt{7}/2$.
\begin{figure}[htbp]
\begin{center}
\includegraphics*[width=10cm, draft=false]{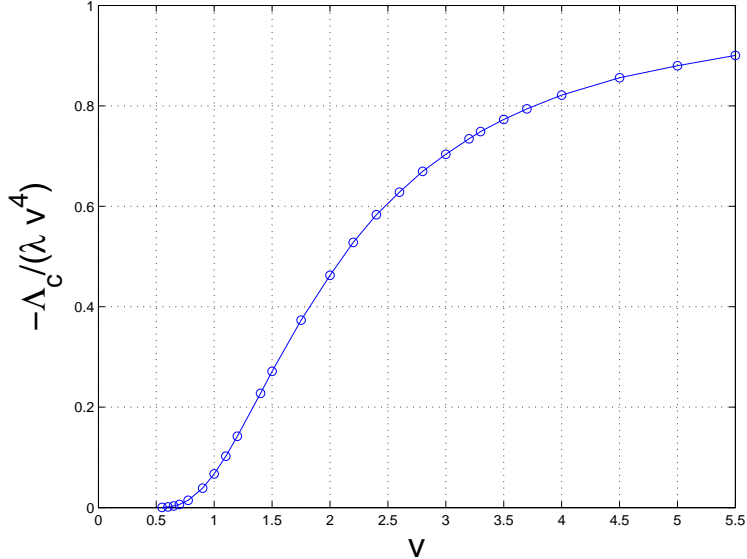}
\caption{Plot of $-\Lambda_c/(\lambda v^4)$ as a function of  $v$.}
\label{fig7}
\end{center}
\end{figure}

Finally, we should comment on the phenomenological applications of our results. One of the 
motivations to study these kind of set-ups is to try to generate a large hierarchy between the
higher dimensional and the $D=4$ Planck masses  in order to explain the so-called hierarchy
problem of traditional grand unified theories. Here these two scales are related by the following
equation
\be
M_{\rm Planck}^2 = \frac{2 \pi M_6^4}{\lambda v^2} \int dx M^2(x) L(x) \;\;,
\ee
where, as explained after eq.~(11), $m(x)=M'(x)/M(x)$. We have checked that, in the solutions
we have found, the hierarchy is never larger than a few orders of magnitude (order $100$ for
$v=1$, increasing to order $1000$ for $v=0.7$), and obtaining a $D=6$ scale of the order of
the TeV proves numerically unachievable. It would correspond to a region with very small $v$, as
already pointed out in Ref.~\cite{Gregory:1999gv}, which we cannot access given that, as mentioned
before, our problem turns into a two-scale one, and our numerical tools become insufficient.

\section{Discussion and Conclusions}

There are in the literature several confining gravity solutions that involve 
scalar vortices in six dimensions \cite{Gherghetta:2000qi,Giovannini:2001hh,deCarlos:2003nq}.
In particular, complete numerical solutions have been presented in two cases \cite{Giovannini:2001hh,deCarlos:2003nq}.  
It is interesting to compare our work with these solutions.
 
In the first one, Ref.~\cite{Giovannini:2001hh}, Giovannini et al. studied a general gravity trapping abelian 
vortex. The presence of the gauge field is crucial and 
determines the properties of the solution. It is described 
by three parameters, $\Lambda$, $v$, and a new one, $\alpha$, that fixes 
the gauge coupling. The authors find 
that the condition of gravity trapping translates into a fine tuning among these three parameters,
fixing a critical surface in this space. The corresponding metric, 
that is a solution to the field equations,
exhibits both Minkowski and angular warp factors that decrease when $r$ increases. This has to be compared to 
our case, where only the Minkowski factor decreases since the solution is cigar-like. This fine tuning is a generic 
fact in this kind of models, and appears when we try to connect the required behaviours for the metric:
a regular one at the origin and a confining one for large values of $r$. The presence of the gauge field does not 
change this fact.

 In the second one, Ref.~\cite{deCarlos:2003nq}, a cigar-like solution based on a BPS scalar vortex was studied. 
The potential was derived from a Supergravity (SUGRA) like theory and its structure was almost fixed. In
particular, there was no room for a cosmological constant. Nevertheless, this SUGRA inspired potential is 
negative near the local minimum, acting like a negative cosmological constant for values of the scalar field 
close to it. The choice of the potential was dictated by simplicity, since for that specific form the field 
equations are first order (for more details see Ref.~\cite{Carroll:1999mu}). Therefore numerical solutions are 
easily obtained just by integrating these first order equations together with regular boundary conditions at the
origin, which happen to be gravity trapping. The SUGRA like structure induces the connection of the two 
(origin and large $r$) required metric behaviours in the solution. Notice that there is an implicit fine tuning 
in the model in the choice of the potential. If we slightly change the coefficients of the terms 
appearing in the potential, 
the new model will almost certainly not 
admit a confining solution. This is related to the 
stability of the solution against radiative corrections, which is an issue is beyond the scope of our work.

To conclude, in this paper we have analyzed the global string in a six-dimensional space time with a negative bulk 
cosmological constant, $\Lambda$. We have presented numerical evidence of the existence of 
solutions that confine gravity. For every value of $v$, the Higgs vev, there is unique value of $\Lambda$ that provides 
a regular solution. This critical cosmological constant is bounded by $- V(0) < \Lambda_c < 0$ and 
approaches its lowest value in the strong gravity limit.
On the other hand, it is difficult  to get a  hierarchy between $M_6$ and 
$M_{Planck}$, at least in the range we have been able to explore numerically.

\acknowledgments
                                                                               
We thank  Ruth Gregory and  Massimo Giovannini for discussing their work in detail. JMM thanks the CERN Theory Division 
for its hospitality. BdC thanks Nuno Antunes, Mark Hindmarsh and Paul Saffin for discussing numerical issues and for 
their very useful advice, and the IEM and IFT (Madrid) for their hospitality.
This work is supported by PPARC (BdC), the Spanish Ministry of Science and
Technology through a MCYT project (FPA2001-1806) 
and by the RTN European Program HPRN-CT-2000-00152 (JMM).

\end{document}